# Retracing Reconstruction

***Establishing an analytical method for the comprehension and systematisation of urban metamorphosis following extreme events***

Mattia Bertin, Università Iuav di Venezia, Department of Architecture and Arts, mbertin@iuav.it

Jacopo Galli, Università Iuav di Venezia, Department of Architecture and Arts, jacopogalli@iuav.it

Francesco Rossi, Università degli Studi di Padova, Dipartimento di Matematica "Tullio Levi-Civita", francesco.rossi@math.unipd.it

### *Abstract*

The problem of identifying models of post-disaster reconstruction is an issue that has been dealt with in depth in a number of works, mostly dedicated to individual cases or to the comparison of models. There are also a number of works that have attempted a systematisation on an historiographic basis. To date, however, there is no overall work aimed at constructing a quantitative method for evaluating and comparing reconstruction experiences. This article proposes to establish a scientific method for the systematic classification of post-disaster reconstructions, based on data analysis. The method consists of four phases: the redrawing of cases with a precise replicable technique; the choice of indicators; the measurement of urban development; the classification of cases. The method is applied to cases that are comparable with each other in terms of type of event and period: we have chosen post-World War II reconstructions in Europe given the larger availability of documentation. The paper leads both to the consolidation of the model as effective and replicable and to the construction of the application, which revises the qualitative classification of most of the addressed cases.

# Retracing Reconstruction

***Establishing an analytical method for the comprehension and systematisation of urban metamorphosis following extreme events***

## Abstract

The problem of identifying models of post-disaster reconstruction is an issue that has been dealt with in depth in a number of works, mostly dedicated to individual cases or to the comparison of models. There are also a number of works that have attempted a systematisation on an historiographic basis. To date, however, there is no overall work aimed at constructing a quantitative method for evaluating and comparing reconstruction experiences. This article proposes to establish a scientific method for the systematic classification of post-disaster reconstructions, based on data analysis. The method consists of four phases: the redrawing of cases with a precise replicable technique; the choice of indicators; the measurement of urban development; the classification of cases. The method is applied to cases that are comparable with each other in terms of type of event and period: we have chosen post-World War II reconstructions in Europe given the larger availability of documentation. The paper leads both to the consolidation of the model as effective and replicable and to the construction of the application, which revises the qualitative classification of most of the addressed cases.

## 1. Introduction

The aim of the paper is to define which urban parameters better describe and allow to critically assess the different processes of urban metamorphosis derived from different design strategies applied in post-disaster reconstruction. A relevant bibliography has been devoted to the subject with a focus shifting from the construction of an historical perspective (Hippler, 2014), to the issue of cultural identity and heritage preservation (Bevan, 2006. Bold et al, 2017. Allais, 2018); from geopolitical and economic reverberations (Coward, 2004. Ikle, 2005) to the use of ICT tools for investigative purposes (Weizman, 2011, 2018), to military tactics adapted to urban planning (Porteous&Smith, 2001. Franke, 2003). In the field of urban studies several researches, mainly devoted to post-WWII Europe, aimed to construct an organised history of reconstruction processes (Mamoli&Trebbi, 1988. Diefendorf, 1990. Cogato-Lanza, 2009. Johnson-Marshall, 2010) exploring the different design approaches in terms of urban and architectural strategies. (Fabietti, Giannino, Sepe, 2013, Lindell, 2013, Shwab et al., 2003, Vale, Campanella, 2005) Nevertheless, a work that aims to identify a scientific approach able to distinguish the different types of urban metamorphoses is still missing.

To conduct this research, we built a team consisting of an architect[1], an urban designer[2], and a mathematician[3]. The architect defined and carried out the critical redrawing of the case studies through the definition of a representation system capable of acting as the basis for the research; the urban designer individuated and described the indicators needed to compare the different reconstruction models, while the mathematician conducted the work of standardizing the analysis of the indicators. The subdivision of tasks prevents the observers from biasing the choice based on preconceived beliefs. In this article we only consider reconstructions after WWII in Europe in order to establish a homogeneous study environment. The article should be considered as the first result

---

of a research that aims to be expanded to different types of disasters (Shi, 2019) and different urban patterns given by geographical, cultural and historical specificities.

For the purpose of this study, we move from a subdivision in categories proposed by Marcello Mamoli and Giorgio Trebbi in their work *L'Europa del Secondo Dopoguerra* (Mamoli&Trebbi, 1988) within the seminal series *Storia dell'Urbanistica* published by Laterza. Despite being published in 1988, the book remains one of the most updated comparative researches in the field of urban design following major disasters. The categories individuated are purely storiographic and based on archival research and personal judgement, while most of the work focuses on the planning rather than on the concrete results. Categories are the following (Figure 1):
- *As it was where it was*: same urban pattern and same building types
- *Continuity between tradition and innovation*: same urban pattern, new building types
- *Rupture with the past*: new urban pattern, same building types
- *Programmatic innovation*: new urban pattern and new building types

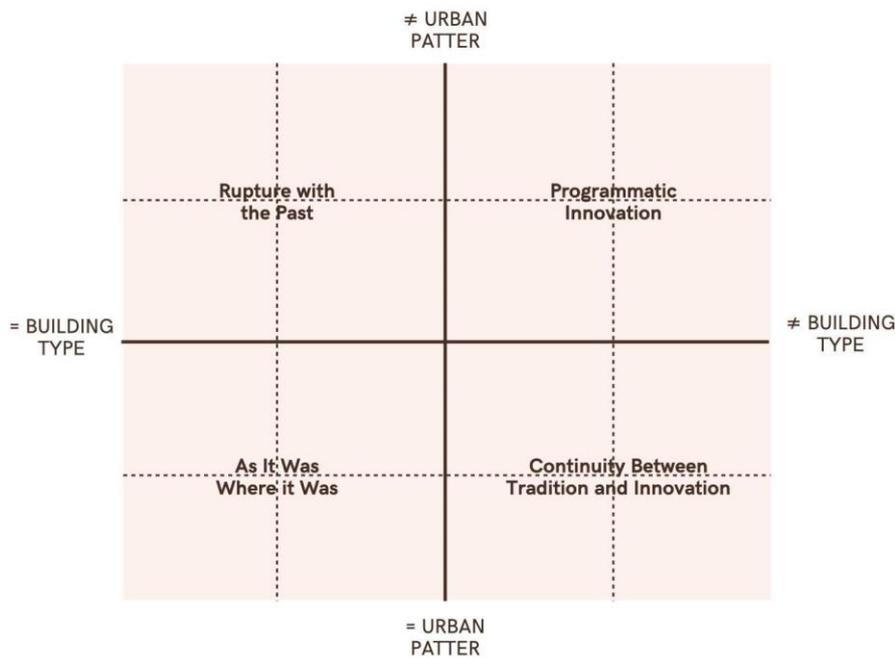

*Figure 1. The four categories of reconstruction distributed in a grid with on the abscissa axis the building type and on the ordinate axis the urban pattern, the used subdivision is an interpretation of Mamoli&Trebbi, 1988. Graphic restitution by \*, here in first publication.*

The book provides examples for each category that acted as the first group of case studies, then extended through further bibliographical and archival research. In the following research steps we will proceed in drafting the same chart, based only on the mathematical analysis of urban indicators. The final aim of the work is to construct and apply a quantitative system based only on the mathematical analyses of urban morphologies; this classification can act as a basis for future urban interventions, informed by a clear dataset in a vision that does not aim to transform urban design in a purely mechanical technical operation but rather to increase the level of knowledge and provide an operational base for designers.

## 2. Theoretical framework
A huge knowledge gap exists in the analysis and systematisation of processes of urban metamorphosis following conflicts, large scale social problems or natural disasters: a comparative study of past cases highlighting similarities and differences, allowing for their critical assessment, has never been fully developed. This paper aims at filling such a knowledge gap, by defining a system of critical redrawing, urban indicators and mathematical analyses that allows to compare different reconstruction strategies in light of the evolution of the urban pattern following these interventions.

*2.1 Critical redrawing framework*

The first research step is carried out through critical redrawing, a fundamental tool to clearly understand the post-disaster reconstruction processes by defining the urban metamorphosis following extreme phenomena. The redrawing system we have adopted is based on updating the methods for the description of urban transformations used, in his books as well as in his urban designs, by the world-renown architectural historian and urban planner Leonardo Benevolo (Albrecht&Magrin, 2015, 2016).

For the purpose of this study we developed an analytical approach adapting what Benevolo defined as *sceneggiatura delle trasformazioni fisiche* (screenplay of physical transformations), where architectural designs or urban environments are described and defined through all the specific characteristics of the object and its context as it would happen in a script for a film or theatrical production. One of the most important scripts written and drawn by Benevolo is the clear illustration of the design processes developed for the San Pietro complex in Rome and presented in Casabella n.572 in 1990 (Benevolo, 1990, 2004). Benevolo's screenplay identifies three key moments in the history of the urban complex through its critical redrawing:

1. The condition of the square before Gianlorenzo Bernini's project,
2. The completion of the colonnade with parallel arms and the definition of the ovoid square by Bernini between 1662 and 1670;
3. The current conditions following the demolition of the Spina dei Borghi and the construction of Via della Conciliazione on a design by Marcello Piacentini and Attilio Spaccarelli completed between 1937 and 1950.

The three phases are not only described and documented, but also drawn at the same scale and with the same type of representation, in order to eliminate the discrepancies given by the different drawing styles. This method of analysis allows us to understand the reasoning behind each design choice that cannot be explained through the simple observation of the current state.

The final drawing proposed in the Casabella article is the cornerstone of the analysis, the transformation map (Figure 2): it superimposes the condition before Piacentini's intervention on the current one and shows with just three layers the complex intertwining of urban continuities and interruptions that have characterized the urban history of the site. The drawing is presented in three simple colors: the red buildings are unchanged in the two periods, the yellow represents the demolitions and the blue the reconstructions, while the dashed yellow and blue shows the buildings that have been rebuilt on the site of the previous buildings. The sum of these temporal layers defines a powerful tool for the understanding of urban metamorphosis. Time becomes a design factor like space and the representation of the different timeframes contributes in a central way to the understanding not only of the evolutionary process but above all of the visible structure, which is only a present concretization of complex phenomena that could, and still can, radically change design choices.

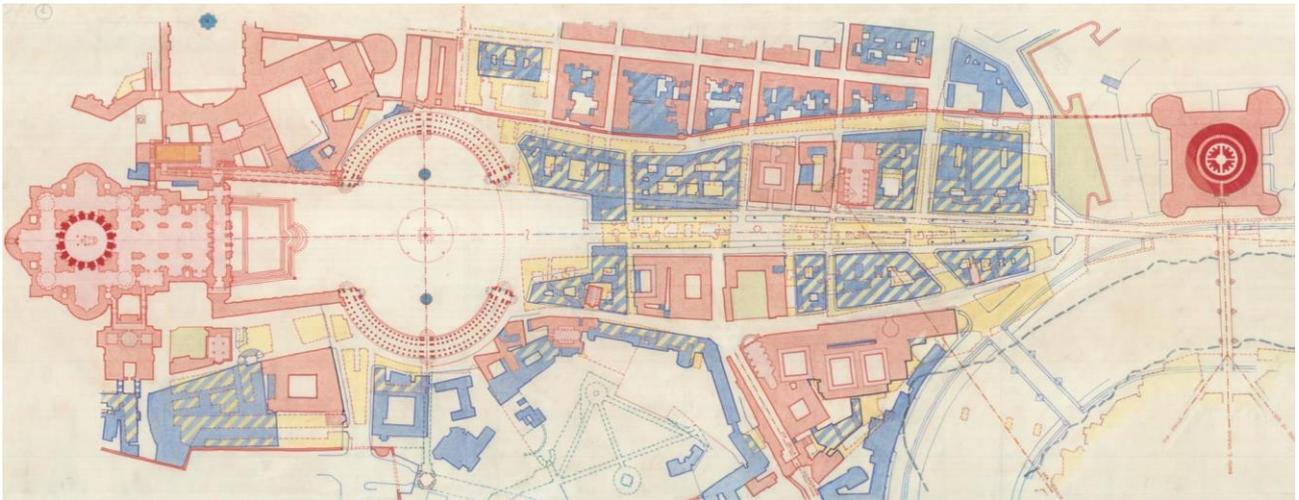

*Figure 2. Transformation map of the Vatican with the overlap of Benini's design and Piacentini's disembowelment. Source: Leonardo Benevolo, La percezione dell'invisibile: piazza san Pietro del Bernini, Casabella n. 572, 1990, pp. 54-60*

## 2.2 Urban indicators framework

To carry out the selection of the indicators able to define the types of reconstruction, we adopted a method that is as much as possible accountable and quantitative. By reviewing the fundamentals of urban planning theory and method, we identified all the possible accounting indicators of permanent characteristics of the built city. The choice of the indicators is strongly indebted to two works in particular: *City sense and city design*, by Kevin Lynch (1995) and *Tecniche Urbanistiche*, by Patrizia Gabellini (2001). In the second volume we have to mention the contributions of Bertrando Bonfantini (Bonfantini, 2007) and Antonio Longo, which enriches the book with definitions and qualitative features on the materials of the open spaces. The two volumes set a canon for the description and reading of urban spaces, declaring the objects and the relationships between objects in a clear and unequivocal way. In the work carried out on the basis of these two references, we removed any element of interpretative acts that might lead us to repeat an individual analysis of reconstructions, since this article is instead devoted to the canonisation of a reading based on the measurement of objects and relations. (Brown et al. 2010) The indicators need to allow a repeatable measurement of built objects and their relationships in the urban environment starting from two-dimensional maps. The elements had to be readable without any confusion by a calculation programme, without the intervention of an operator describing the objects map by map. The indicators selected to describe the reconstructions and analyse the case studies were 18. They have been labelled as *Starting Indicators (SI)* and numbered from 1 to 18. Some are direct indicators and others are secondary indicators, the result of comparing the former. The list of indicators considered is as follows:

- *SI1: Destroyed surface on built-up area;*
- *SI2: Previous occupied surface;*
- *SI3: Number of elements previous;*
- *SI4: Median size of elements previous;*
- *SI5: Average distance between elements previous;*
- *SI6: Surface squares previous;*
- *SI7: Surface occupied next;*
- *SI8: Number of elements next;*
- *SI9: Median size of elements next;*
- *SI10: Average distances of elements next;*
- *SI11: Area of squares next;*
- *SI12: Area occupied next over previous;*
- *SI13: Number of elements next over previous;*

- *SI14: Median size of elements next over previous;*
- *SI15: Average distances of elements next over previous;*
- *SI16: Area of squares next over previous;*
- *SI17: Site maintenance;*
- *SI18: Street level maintenance.*

This list of indicators allowed us to develop an initial broad analysis of the relationship between the different case studies. These indicators can be grouped in three classes: in a first class, indicators describe the architectural space before the destruction (SI2-SI6); in a second class, the same indicators are computed after the destruction (SI7-SI11). The third class is the comparison of urban characteristics before and after the destruction; this class can be seen as an indicator of the transformation of the urban space (SI1 and SI12-SI18). It is important to remark that these indicators are all quantitative and can directly be computed through the analyses of produced drawings. In this sense, they can be completely computed by objective, available data and do not suffer from personal judgment. Finally, observe that no data about the 3D structure of the city is present in the indicators, due to the lack of reliable information of this kind before the reconstruction.

### 2.3. Mathematical analysis

The key idea behind the mathematical analysis of the chosen indicators is to develop an objective classification of urban metamorphosis, driven only by the data. Moreover, the goal of finding synthetic and comprehensible indicators is also crucial: this idea is also coherent with the aim of future application of the method to provide comparable base data for the design of urban transformations in response to extreme events.

Both objectives are met by choosing the "sparse feature selection" method proposed by (Witten&Tibshirani, 2010): given a dataset with some quantitative features, it both selects the most relevant features and provides their relative weights. The method is sparse, in the sense that it aims to select a small number of features, in coherence with our aim of synthetic indicators. Moreover, the selection of features, instead of the creation of new features based on mixing the original ones (as, e.g., with Principal Component Analysis), allows us to keep comprehensible selected features.

### 3. Materials

The transformation map tool identified by Benevolo was produced by the research team for a series of post-WWII reconstruction case studies. The operation requires a considerable amount of work in order to identify the cartographic materials or historical photographs that document the state before the destruction, the level of destruction and the reconstruction process. The process of standardization of the different forms of representation, conducted through a careful redrawing allows to eliminate the relevant discrepancies (scale, type of measurements, level of detail) and achieve a full overlap of the different timeframe maps.

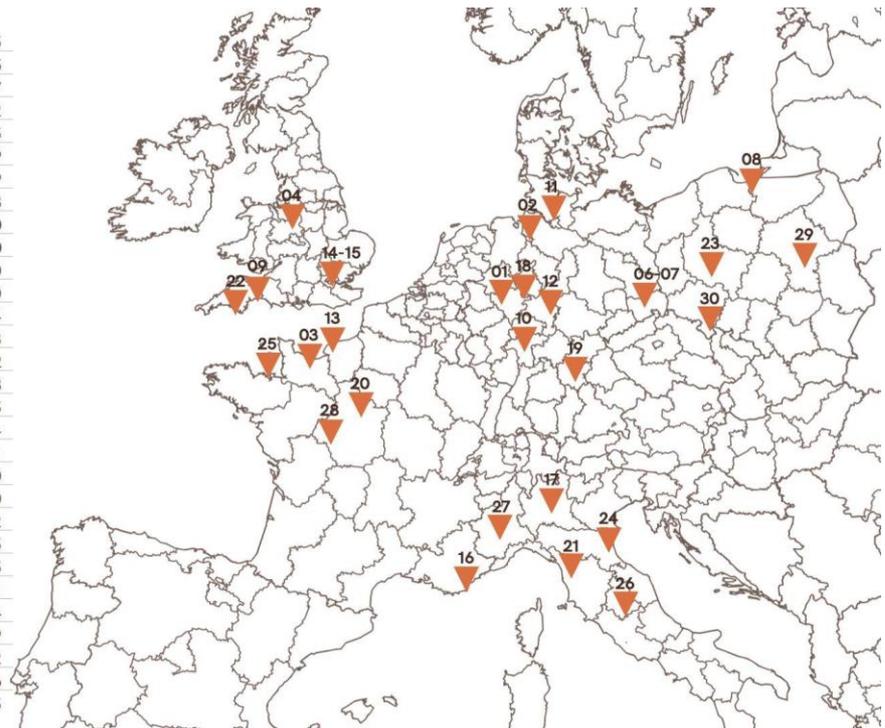

| 01 | Bochum | GER | 1943 | 1965 |
| 02 | Bremen | GER | 1942 | 1955 |
| 03 | Caen | FRA | 1944 | 1957 |
| 04 | Coventry | GBR | 1943 | 1962 |
| 05 | Den Haag [Vogelwijk] | NLD | 1941 | 1965 |
| 06 | Dresden [Altstadt] | GER | 1945 | 1989 |
| 07 | Dresden [Seevorstadt] | GER | 1945 | 1989 |
| 08 | Elblag | POL | 1939 | 1983 |
| 09 | Exeter | GBR | 1942 | 1950 |
| 10 | Frankfurt Am Main | GER | 1944 | 2010 |
| 11 | Hamburg | GER | 1943 | 1960 |
| 12 | Kassel | GER | 1943 | 1970 |
| 13 | Le Havre | FRA | 1944 | 1964 |
| 14 | London [Bankside] | GBR | 1940 | 1968 |
| 15 | London [Barbican] | GBR | 1940 | 1982 |
| 16 | Marseille [Vieux Port] | FRA | 1943 | 1958 |
| 17 | Milano | ITA | 1944 | 1965 |
| 18 | Münster | GER | 1944 | 1964 |
| 19 | Nürnberg | GER | 1945 | 1971 |
| 20 | Orleans | FRA | 1940 | 1960 |
| 21 | Pisa | ITA | 1943 | 1960 |
| 22 | Plymouth | GBR | 1940 | 1962 |
| 23 | Poznan | POL | 1939 | 1965 |
| 24 | Rimini | ITA | 1943 | 1965 |
| 25 | Saint Malo | FRA | 1944 | 1961 |
| 26 | Terni | ITA | 1943 | 1954 |
| 27 | Torino | ITA | 1943 | 1959 |
| 28 | Tours | FRA | 1940 | 1962 |
| 29 | Warsaw [Muranow] | POL | 1939 | 1956 |
| 30 | Wroclaw | POL | 1939 | 1965 |

*Figure 3. List and map of the geographical distribution of the selected case studies: Bochum (1943-1965), Bremen (1942-1955), Caen (1944-1957), Coventry (1943-1962), Den haag (1941-1965), Dresden (1945-1989), Eblag (1939-1983), Exeter (1942-1950), Frankfurt Am MainFrankfurt Am Main (1944-2010), Hamburg (1943-1960), Kassel (1943-1970), Le Havre (1944-1964), London (1940-1982), Marseille (1943-1958), Milano (1944-1965), Münster (1944-1964), Nürnberg (1945-1971), Orleans (1940-1960), Pisa (1943-1960), Plymouth (1940-1962), Poznań (1939-1965), Rimini (1943-1965), Saint Malo (1944-1961), Terni (1943-1954), Torino (1943-1959), Tours (1940-1962), Warsaw (1939-1956), Wroclaw (1939-1965).*

The examined case studies are distributed among France (6), Germany (9), Great Britain (5), the Netherlands (1), Italy (5) and Poland (4). Geographical distribution does not fully show the amount of conflict-induced urban destructions, but rather shows the greater amount of archival information available in some contexts. To make the data comparable, one or more frames of 1kmx1km with uniform urban fabric were identified and redrawn as an analyses area, the frames were selected to fully include the area with the highest level of destruction.

The timeframes for the end of the reconstruction processes, which allows the draft of the transformation maps, vary significantly: in some cases, characterized by destruction of a significant level in a punctual area, the reconstruction processes are already completed in the 1950s. (Marseille, Caen, Florence, etc.); most of the planned large-scale reconstructions are completed in the 1960s (Le Havre, Rotterdam, etc.) while in some examples for economic or political reasons the processes lasts until the 1970s (Kassel, etc.) or even to the '80s (Dresden, London, etc.) up to the point of multiple reconstructions that extend well into the 2000's (Frankfurt). A formal date of completion of the reconstruction exists only in some cases of top-down planning such as in Le Havre (Etienne-Steiner, 2018), where in 1964 Auguste Perret's plan was declared complete, or in Rotterdam (Blom et al, 2017) where Cornelius Van Tra's Basis Plan remains in operation until 1968. In many cases the completion of a symbolic building is used as the beginning or end of the reconstruction: in Milan (Pertot&Ramella, 2016) the rapid reconstruction of the La Scala theater with the inaugural concert directed by Arturo Toscanini on 11 May 1946 and the completion of BBPR's Velasca tower in 1957; in Marseille (Bedarida, 2012), the inauguration in 1954 of the new Vieux Port district which was deliberately razed during the Nazi occupation and rebuilt on Fernand Pouillon's design; or in London (Marmaras, 2014) where the Cripplegate area sees a radical upheaval of urban forms which culminates with the inauguratio of the Barbican Center in 1982. The study of available historical research on each case study and of comparative works allowed us to place each case study in the synoptic chart, solely on the basis of qualitative attribution informed by acquired historical knowledge.

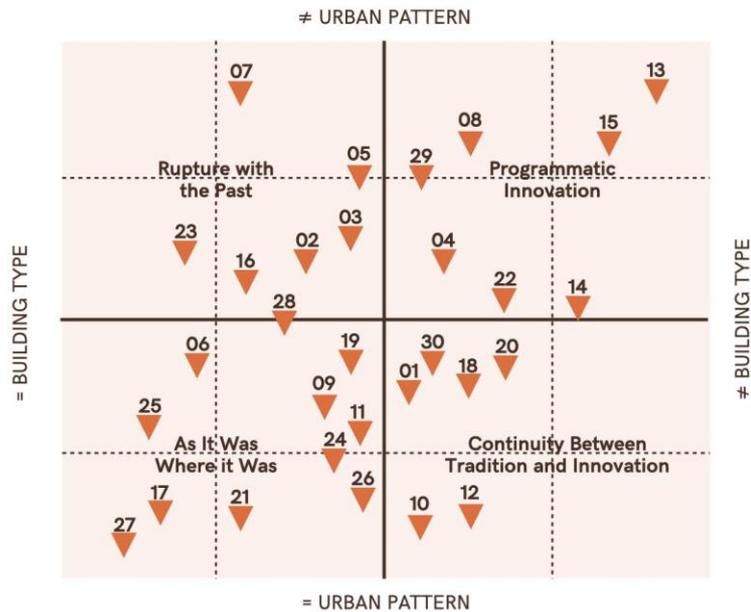

*Figure 4. Qualitative classification of case studies based on bibliographic research.*

## 4. Methods

We carried out the research in 4 steps: critical redrawing; measurement of indicators; weighting and selection of indicators; alteration of the synoptic chart. The first step consists in the redrawing of 40 case studies in a coherent and comparable way. Four drawings were produced for each city: before destruction; after destruction; reconstruction; transformation map. The first three drawings crystallise in a single drawing a timeframe and become the tool for metrically analysing the cases, and redefining the types of reconstruction. This phase, described in detail below, has allowed the development of a homogeneous and comparable catalogue of examples. After this phase, we excluded those cases in which the drawing was not perfectly comparable, or in which non-significant but potentially unrecognisable discrepancies could appear. This resulted in 30 cases distributed throughout Europe. Once we had four drawings per city, we measured the data for the selected indicators. The calculation programme carried out a restitution in table form of each of the chosen indicators based on the analyses of each transformation map. At this point we carried out an automated selection of the most meaningful indicators and their relative weight. The weighting between indicators is purely statistical, based on the data themselves, in order to avoid biasing by the observers' interpretation. The outcome of the third step was the choice of most meaningful indicators and their relative weight.

In the fourth step, we repositioned the case studies with respect to what is recognised in the literature. This distribution is a direct result of the quantitative analysis, since it is based on computation of transformation scores for both urban and building indicators. It allows to comment on the mismatch between the perception in the reading of each case in history and its actual location. It is an important descriptor of the relationship between what has been proposed so far in the reading of reconstruction models and a scientific restatement of these cases.

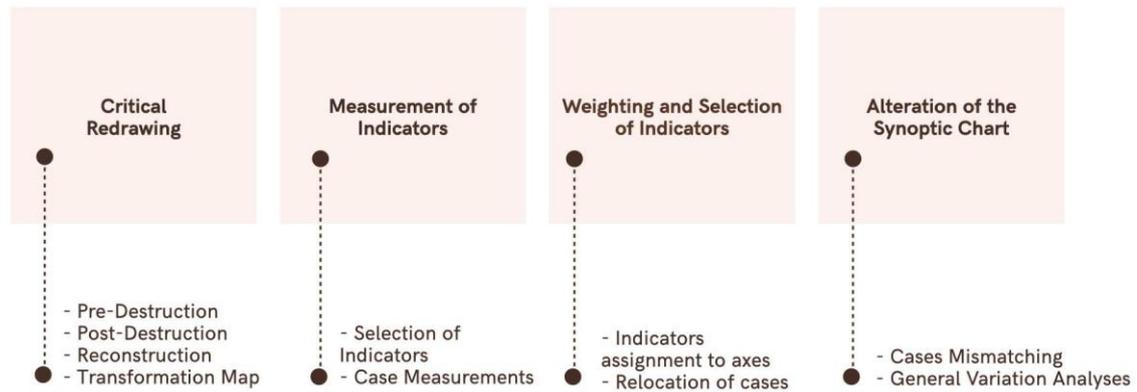

*Figure 5. Methodological flowchart: description of the research development and elaboration process. Here in first publication.*

## 4.1. Critical redrawing

The critical redrawing system is devoted to the construction of a set of drawings, all drafted with the same graphic approach, which allows to clearly understand the sequence of interventions and the process of morphological metamorphosis that led to the formation of the current urban contexts. An operation, never carried out in a comparative way, which makes it possible to re-evaluate the numerous monographic studies on post-war reconstruction processes (Cohen, 2011. Düwel&Gutschow, 2013. Moravánszky, 2016) in light of a common evaluation system and therefore to define interpretative categories that are not limited to the observation of archival materials but are informed through the tools of redrawing and its interpretation. Retracing the transformations allows not only to testify past experiences, but it builds an operational tool for the understanding of current urban dynamics, with the ultimate goal of orienting future intervention strategies. The redrawing allows to illustrate the three different urban structures before, during and after the extreme events and is concluded, as in Benevolo's example, with the drafting of a transformation map that overlaps the three moments and clearly shows the process of urban metamorphosis. The redrawing then produces the following:

- *D1: Pre-Destruction Map;*
- *D2: Post-Destruction Map;*
- *D3: Reconstruction Map;*
- *D4: Transformation Map.*

The systematic drafting of transformation maps makes it possible to compare the spatial consequences on the urban pattern of the various reconstruction strategies applied. The result frequently shows a clear difference with respect to storiographic publications which, focusing on the formal aspects of the plans rather than on their practical results, tends to re-propose simplified urban forms that often are not fully confirmed in reality.

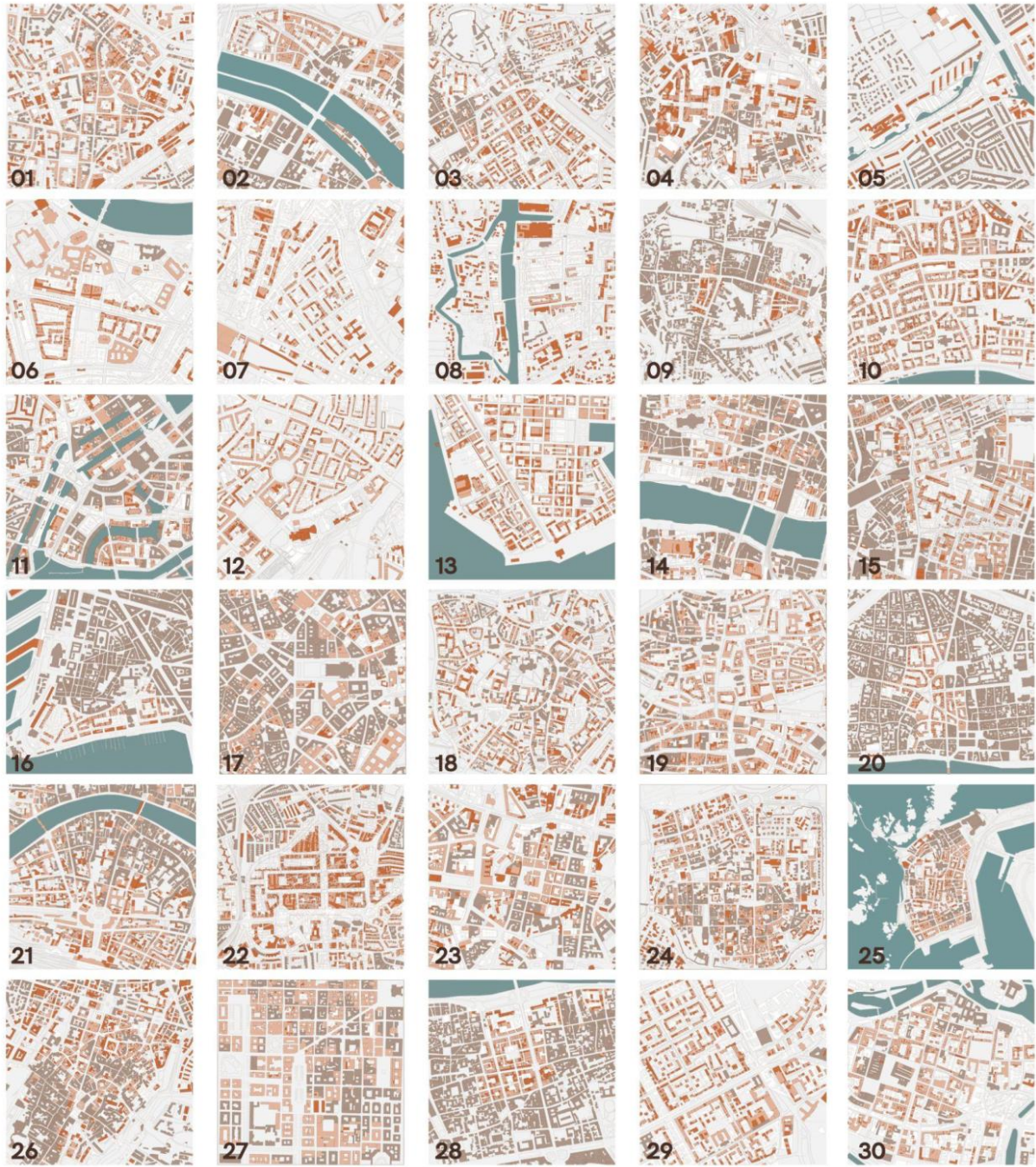

*Figure 6. Transformation maps drafted for the 30 case studies: in brown unvaried, in orange reconstructed in white destroyed and not reconstructed, hatched reconstructed over destroyed.*

## 4.2. Measurement of indicators

The second step of the project is the computation of the selected indicators for the case studies, based on the drawings produced in the previous step. Given the three drawings (D1-D3), it is possible to directly compute each indicator with standard imaging tools. In our case, we used standard MATLAB routines (such as the ones in the Image Processing Toolbox) to identify buildings, squares and their perimeters in an automated manner. With this method, it was possible to compute all indicators, as follows:

- From a comparison of D2 over D1, it is possible to recover I1;
- From D1, it is possible to recover I2-I6;

- From D3, it is possible to recover I7-I11;
- From I2-I11 computed above, it is possible to recover I12-116;
- The comparison of D1-D2-D3 allows us to compute I17-I18.

### 4.3. Weighting and selection of the indicators

The third step of our classification is a statistical analysis of the indicators computed above, with the aim of selecting the most relevant and weighting them with the aim of classifying the case studies. In this project, this step is certainly the one in which mathematical/statistical tools play the most crucial role, based on the method of feature selection proposed by (Witten&Tibshirani, 2010) recalled above. During this step, we carried out a significance analysis of each indicator. This allowed us to eliminate indicators that were underrepresented in terms of significance. In the next steps of the investigation, we kept only those indicators that were significant to the evaluation. The new indicators have been labelled as Indicators (I) and numbered 1-8.

After normalizing each indicator, we collect them in two groups: Indicators of transformation of the urban pattern (U) and Indicators of the transformation of building type (B). The (U) indicator is the weighted sum of indicators in this group, where the weight is given by the feature selection process above. The same holds for the (B) indicator. Due to the necessary normalization of the original indicators, the key information of the indicators (U-B) is in the comparison between different cases and not their actual value.

### 4.4 Alteration of the Synoptic Chart

The last computational step led us to redistribute the analysed cases on the new grid. This grid is the result of the measurement with the selected indicators. Each case is positioned on the two axes according to the values expressed by the indicators. For each case we established an initial position, that of literature, and a final position, that of measurement. The distribution of cases is not intended as a precise objective position, but rather as a domain of belonging. The instrument is at its first test and is not yet consolidated to make detailed metric evaluations. On the other hand, the final domain of each case is very clear and objective. This step will also allow us to delineate a distribution cloud of cases, which will be useful to understand the general behaviour on reconstructions, if any. Furthermore, this cloud will allow us to define if there is a relationship between architectural innovation and urban innovation, or if the two axes are independent.

## 5. Results

The progress of the research has led to a stable formulation of recognised indicators and related classes of belonging. Firstly, we describe how we selected the significant indicators and how these indicators were assessed in their correlation. Secondly, we report the measure of each case for each indicator. Thirdly, we report the distribution of the cases in a new grid, composed by the selected indicators.

### 5.1. Primary and auxiliary indicators

A first weighting of the whole 18 indicators shows that indicators related to map transformation (SI12-SI18) are among the most relevant, with a cumulative weight of more than 63%. Moreover, the other most relevant index SI4 (with weight 26%) is strongly correlated to both indexes SI13 (Pearson Correlation coefficient 0.75) and SI14 (PCC 0.64). For this reason, we decide to further restrict our indicators to the ones describing map transformation, i.e. SI1 and SI12-I18. The result is this list of indicators mathematically recognised as relevant:

- *I1: Destroyed surface on built-up area*
- *I2: Area occupied next over previous*

- I3: Number of elements next over previous
- I4: Median size of elements next over previous
- I5: Average distances of elements next over previous
- I6: Area of squares next over previous
- I7: Site maintenance
- I8: Street level maintenance

We then classify these indicators as describing either the transformation of the urban pattern (I1, I2, I3, I4, I8) and the transformation of the building type (I5, I6, I7). Given the relative weight of such indicators, for each case study we compute a score of transformation of both the urban pattern and the building type.

| Indicators of transformation of the urban pattern (U) | Indicators of the transformation of building type (B) |
|---|---|
| I1: Destroyed surface on built-up area | I6: Number of elements next over previous |
| I2: Area occupied next over previous | I7: Median size of elements next over previous |
| I3: Area of squares next over previous | I8: Average distances of elements next over previous |
| I4: Site maintenance | |
| I5: Street level maintenance | |

*Table 1: Subdivision of the final indicators into the categories Urban Pattern and Building Type..*

## 5.2. Case measurement

After carrying out the case selection and the redrawing procedure, we measured each case with the eight selected indicators. We report here in the table the result of the measurement. <mark>All indicators are in percentage form.</mark> We report the indicators in the rows in order to facilitate the reading of the mobility of the measures per indicator. In our opinion it is more meaningful here to underline how each indicator differs in the cases than a case by case reading. It is however possible to get a picture of the measures case by case by reading the columns of the table vertically.

| Ind | 1 | 2 | 3 | 4 | 5 | 6 | 7 | 8 | 9 | 10 | 11 | 12 | 13 | 14 | 15 | 16 | 17 | 18 | 19 | 20 | 21 | 22 | 23 | 24 | 25 | 26 | 27 | 28 | 29 | 30 |
|---|---|---|---|---|---|---|---|---|---|---|---|---|---|---|---|---|---|---|---|---|---|---|---|---|---|---|---|---|---|---|
| I1(U) | 76 | 64 | 72 | 69 | 36 | 97 | 100 | 96 | 23 | 92 | 57 | 99 | 98 | 53 | 63 | 26 | 39 | 74 | 79 | 20 | 47 | 80 | 63 | 60 | 63 | 41 | 55 | 26 | 94 | 80 |
| I2(U) | 89 | 70 | 69 | 76 | 93 | 45 | 44 | 82 | 90 | 74 | 75 | 54 | 61 | 75 | 68 | 88 | 92 | 76 | 73 | 92 | 109 | 74 | 85 | 102 | 86 | 109 | 99 | 93 | 51 | 54 |
| I3(B) | 61 | 107 | 115 | 85 | 95 | 59 | 69 | 61 | 105 | 97 | 99 | 78 | 108 | 88 | 112 | 78 | 116 | 60 | 100 | 104 | 94 | 61 | 97 | 139 | 99 | 100 | 104 | 111 | 339 | 83 |
| I4(B) | 134 | 83 | 51 | 88 | 81 | 147 | 30 | 122 | 72 | 68 | 84 | 60 | 54 | 85 | 69 | 128 | 82 | 107 | 71 | 87 | 100 | 107 | 79 | 79 | 84 | 84 | 95 | 80 | 17 | 65 |
| I5(B) | 115 | 130 | 125 | 123 | 104 | 122 | 132 | 114 | 100 | 137 | 133 | 152 | 131 | 120 | 164 | 115 | 111 | 135 | 122 | 114 | 88 | 160 | 103 | 102 | 101 | 96 | 101 | 106 | 138 | 142 |
| I6(U) | 223 | 172 | 150 | 266 | 131 | 252 | 224 | 132 | 113 | 129 | 224 | 148 | 155 | 682 | 239 | 266 | 129 | 152 | 159 | 115 | 79 | 322 | 133 | 74 | 102 | 78 | 96 | 108 | 335 | 432 |
| I7(U) | 75 | 92 | 71 | 70 | 76 | 81 | 52 | 27 | 95 | 72 | 90 | 72 | 65 | 94 | 90 | 92 | 97 | 78 | 86 | 97 | 83 | 65 | 80 | 71 | 87 | 77 | 98 | 93 | 70 | 93 |
| I8(U) | 78 | 76 | 67 | 70 | 78 | 51 | 31 | 27 | 95 | 56 | 74 | 52 | 41 | 80 | 71 | 89 | 90 | 80 | 64 | 92 | 80 | 71 | 64 | 65 | 77 | 73 | 96 | 86 | 30 | 68 |

*Table 2: outcome of case measurement through the 8 chosen indicators.*

## 5.3. Reporting of cases in the four domains.

We then displayed the distribution of cases in the generated grid. The cases, measured above, were disposed along the axes as described. We show in the figure below the outcome of the distribution, in addition to the distribution found in the literature and the list of cases to facilitate a synoptic reading.

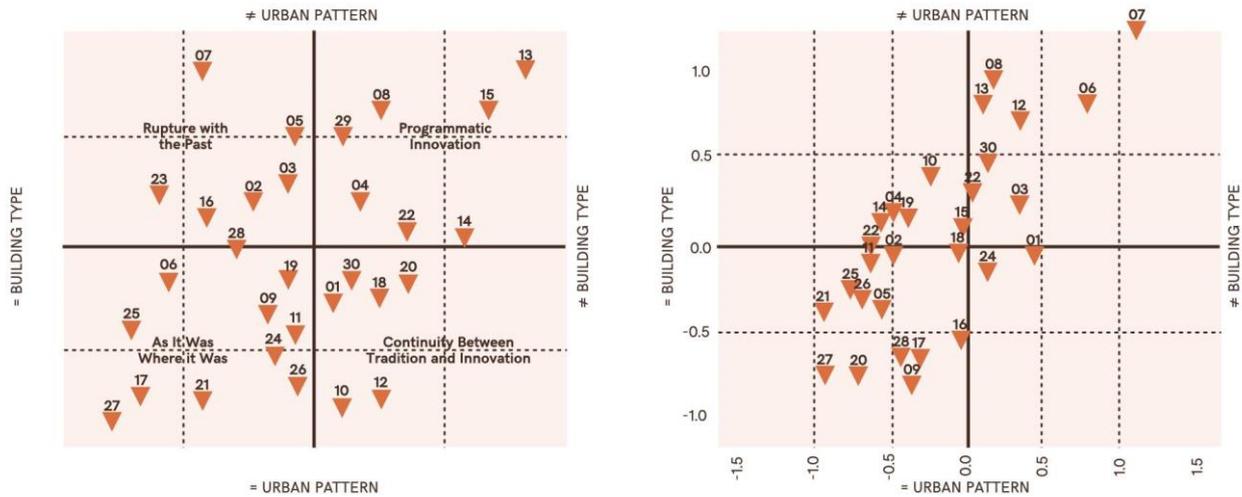

*Figure 7. Alteration of the scheme from the qualitative evaluation (left) to the quantitative one (right).*

Any comment on the outcome is left to the Discussion paragraph, but it is already possible to determine a strong significance of the research for the evident variation of the distribution and of the membership of the stable classes. Below we find two examples of the application of the method. One that led to a confirmation of the location in the grid (Elblag) and one that led to a modification of the location (Rimini). The case of Elblag, top row in Figure 8, shows us a case of major urban and architectural innovation. The case of Rimini, bottom row, in the literature apparently seemed to be a case of only architectural innovation, while in reality it is a case of coherent reconstruction both on the urban and architectural axis.

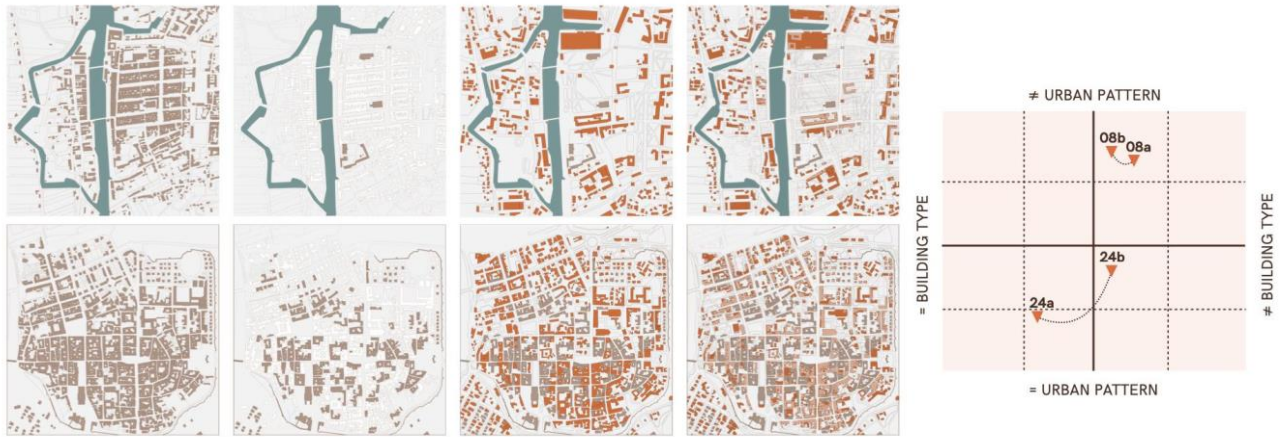

*Figure 8. Drawings for 08 Elblag and 24 Rimini (before event, destruction, reconstruction, transformation) and alteration of the position in the synoptic chart in the qualitative and quantitative system.*

## 6. Discussion

As can be noticed from the comparison of the two diagrams in Fig. 8, the research aims to establish a turning point in post-disaster reconstruction studies. Each step was carried out with the greatest possible clarity and rigour by the research team in order to favour comparative studies or the repetition of the calculation. The first outcome we can suggest here is a scientific method for the evaluation of a post-disaster reconstruction case, starting from the design method to the calculation method, to the classification method. The few final indicators chosen and verified in their significance aim to produce coherent measures in the evaluation of reconstruction processes. The final grid aims to become a working and comparison table for anyone who wants to compare different reconstruction cases.

We are aware of the fact that we did not explore the characters of effectiveness and efficiency of the reconstructions with this method, but it was a precise choice. It would have been misleading to unify a classification of typologies with an analysis of outcomes: we would have had to carry out a process of indicator selection, qualification and modelling also in these aspects. The relation of the indicators could complicate the reading of the results and the repetition of the study. For this reason we are leaving it open to a subsequent study on the effectiveness and efficiency of the reconstructions included in the stable classes identified, either by ourselves or by other teams.

The first evidence we can deduce from the outcome of the research is the substantial placement of many cases. Often the quadrants of the grid contain predominantly different cases in the two representations. This means that the attribution in the literature of the character of reconstruction is often inaccurate when tested by measurement.

The second remarkable aspect concerns the shape of the cloud of points distributed on the grid. We can recognise an important difference with the literature-based cloud. The cloud generated by the literature descriptions of the reconstruction cases tended to occupy the whole grid, without polarisation. The calculation of the measures led to a much less expanded distribution over the areas of the grid where architectural innovation and urbanism do not proceed together. The cloud is strongly polarised on a strong correlation between the degree of urban innovation and of the architectural innovation. The emerging cloud is a first visualisation of the distribution of post-war reconstructive choices in Europe in the second half of the 20th century. The particular distribution assumed by the cloud leads us to ask whether this distribution is a characteristic of reconstructive choices, or whether it depends on other factors. Is it possible that the analysis of a significant sample of reconstructions in North America, or in the Middle East, leads to the same shape? Is it possible that analysis of a sample of post-earthquake reconstructions in Europe would lead to the same shape? In essence: does there exist different cultures of urban and architectural design or is this diagonal shape innate to the ways of human reconstruction?

The research takes the form of an initial approach to reviewing and standardising the practices of reflection on post-disaster reconstructions. The questions posed here open up a broad field of enquiry, and the post-WWII reconstructions themselves discussed here are a sample that needs to be further integrated to confirm this initial expression.

## 7. Bullet Point Summary of Outcomes

- The research tried to measure the placement in stable classes of post-World War II European reconstructions considering the characters of urban and architectural innovation.
- The research used Leonardo Benevolo's pre-destruction, post-destruction, reconstruction redrawing method and standardised the approach.
- The indicators were chosen with a mathematical significance study.
- Most relevant indicators of transformation of the urban pattern: destroyed surface on built-up area; area occupied next over previous; area of squares next over previous; site maintenance; street level maintenance.
- Most relevant indicators of the transformation of building type: number of elements next over previous; median size of elements next over previous; average distances of elements next over previous.
- The distribution cloud of points on the axes of urban and architectural innovation is very different between description in the literature and measurement with indicators.
- There is a strong correlation in the cases treated between grade of urban innovation and grade of architectural innovation in reconstruction.

## 8. Conclusions

The research has led to three main results: a methodical application of Leonardo Benevolo's redrawing method as an approach to the study of post-disaster reconstructions; a selection of the most effective indicators for evaluating reconstruction cases; a recognition of a strong relationship between urban maintenance-innovation and architectural maintenance-innovation. The research is still based on a limited number of cases, but it has above all the value of having woven a replicable method. The article does not consider the issues of effectiveness and efficiency of the reconstructions treated in order not to get into complexities that are difficult to handle, and this remains an open point on which to make subsequent evaluations. In our opinion, the three elements of methodological innovation and the research questions described here allow for a point of collective reflection in order to trace a new path of reflection on the ways and forms of reconstruction.